\documentclass[fleqn,12pt,twoside]{article}
\usepackage[headings]{espcrc1}
\usepackage{graphicx}

\title{Probing Cold and Hot, Dense Nuclear Media Via High-p$_{T}$ Jets
with Di-hadron and $\gamma$-hadron Correlations at PHENIX}

\author{N. Grau\address{Department of Physics \& Astronomy, Iowa State University \\
        Ames, IA 50011} (For the PHENIX Collaboration)
        \thanks{For the full list of PHENIX authors and acknowledgements, see Appendix `Collaboration` of this volume.}}

\runtitle{Probing Cold and Hot, Dense Nuclear Media with Jets}
\runauthor{N. Grau}

\begin{document}

\maketitle

\begin{abstract}
With the recent high statistics Au+Au and Cu+Cu runs at RHIC, it
has become possible to systematically study jet properties in
several different colliding systems with potentially different
final state interactions.  In this talk we present results from
high-p$_{T}$ di-hadron and $\gamma$-hadron correlations from p+p,
d+Au, Cu+Cu, and Au+Au collisions where jets in vacuum, cold
nuclear matter, and hot, dense nuclear matter are studied.
\end{abstract}

\section{Introduction}
One of the most celebrated results from the first three years of
RHIC is the suppression of single high-p$_{T}$ particle production
in central Au+Au collisions due to jet
suppression~\cite{PhenixSuppress}. The effect of jet suppression
was dramatically seen in two-particle correlations where the
away-side jet in central Au+Au collisions was largely
extinguished~\cite{StarAwayJet} and its shape
modified~\cite{PhenixAwayJet}.  These observations are consistent
with calculations from parton energy loss via gluon bremsstrahlung
when the parton traverses a medium with a large gluon density of
$dN_{g}/dy \sim$ 1000~\cite{GVEloss}. Further, the same single
particle suppression and away-side jet suppression is not present
in d+Au collisions~\cite{dAuEnhance} meaning that it is the final
state interactions in the central Au+Au collisions that suppress
the jets. With the recent high-statistics Au+Au and Cu+Cu runs it
is becoming possible to do more detailed studies of the away-side
jet modification with two-particle correlations.

Two-particle correlations are measured in the PHENIX central
spectrometer arms and are defined as
\begin{equation}\label{eqn:corrfcn}
C\left(\Delta\phi\right) \propto
\frac{dN_{real}/d\Delta\phi}{dN_{mix}/d\Delta\phi} \propto
\frac{1}{N_{trig}}\frac{dN}{d\Delta\phi}
\end{equation}
where $dN_{real}/d\Delta\phi$ are the event pair distribution and
the $dN_{mix}/d\Delta\phi$ are the mixed event distributions where
each particle of the pair is chosen from a random event.  The
mixed event distributions correct for the non-uniform pair
acceptance of the PHENIX central arms. These correlation functions
can be turned into a pair per trigger distribution,
$1/N_{trig}dN/d\Delta\phi$ where the normalization is dependent on
the acceptance and the efficiency~\cite{jet:method}. Two particle
correlations exhibit two peaks at $\Delta\phi$=0 (near) and
$\Delta\phi$=$\pi$(away) which represent correlations within the
same jet and between di-jets respectively. In A+A collisions
elliptic flow is also present which causes a harmonic modulation
of the isotropic background.

\section{Jets in p+p and d+Au Collisions}
\begin{figure}[tb]
\begin{minipage}[b]{80mm}
\begin{center}
\includegraphics[width=80mm,height=80mm]{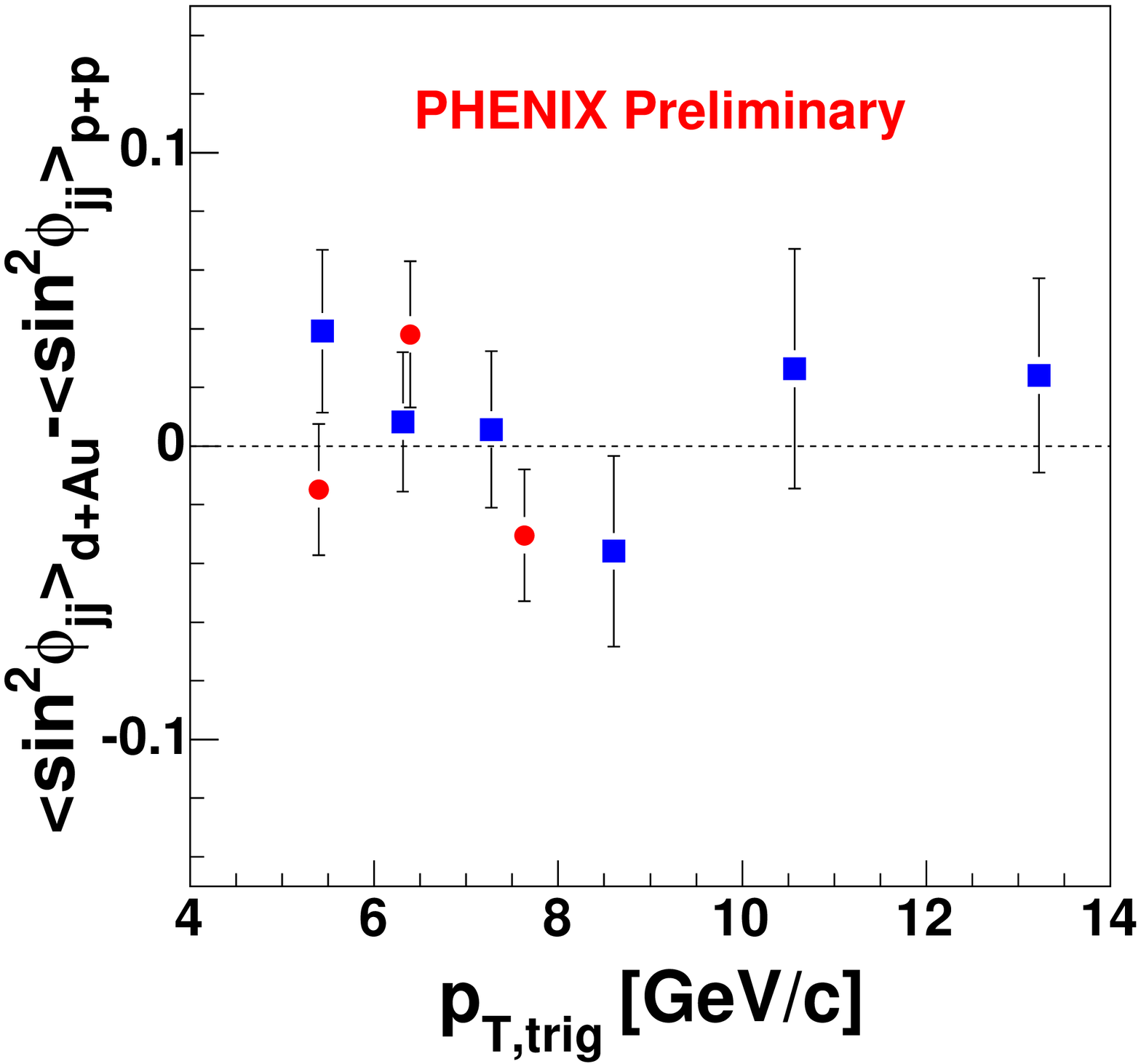}
\caption{Difference of $\rm{sin}^{2}\phi_{jj}$ between p+p and
d+Au.}\label{fig:sinsq}
\end{center}
\end{minipage}
\hspace{\fill}
\begin{minipage}[b]{80mm}
\begin{center}
\includegraphics[width=80mm,height=40mm]{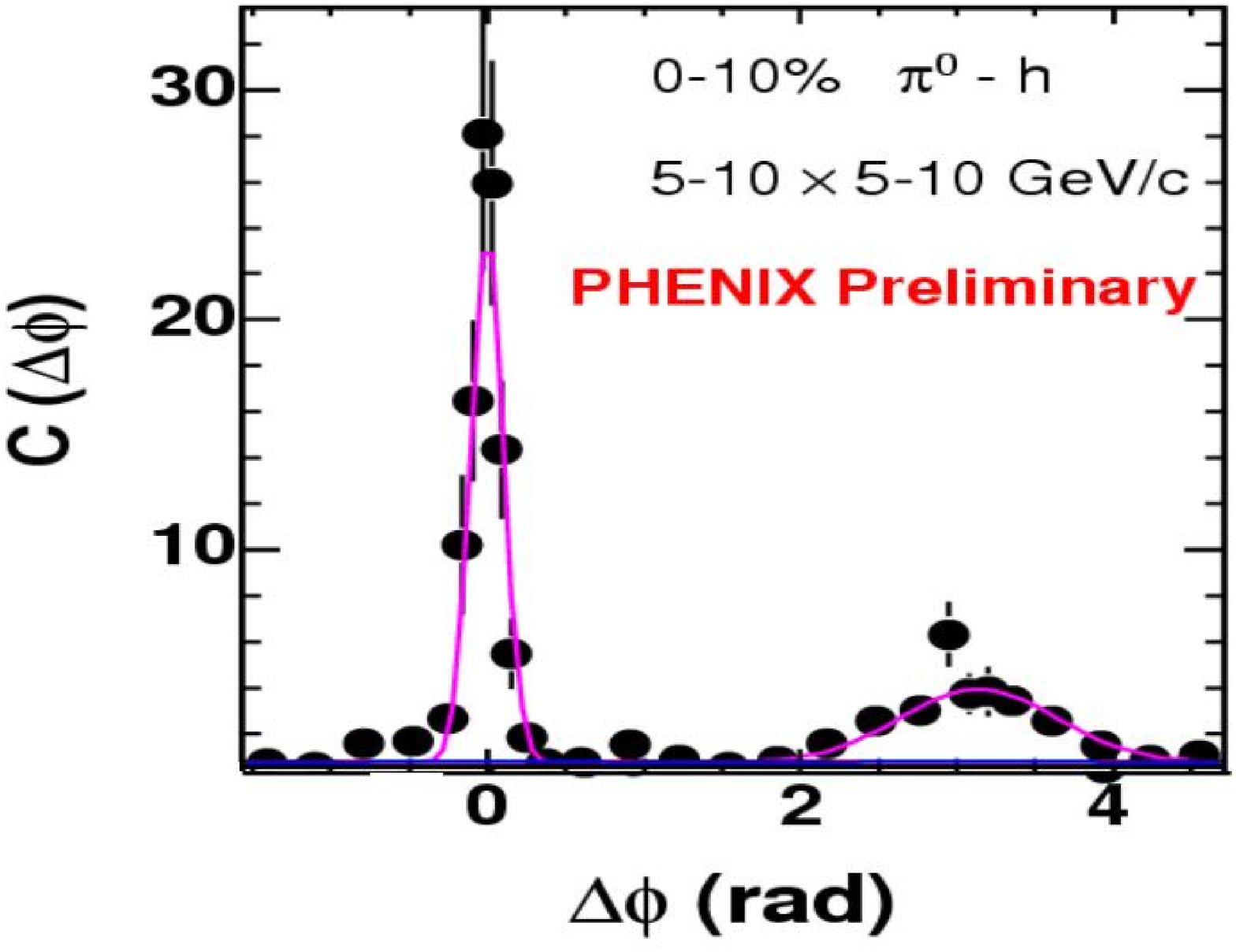}
\includegraphics[width=80mm,height=40mm]{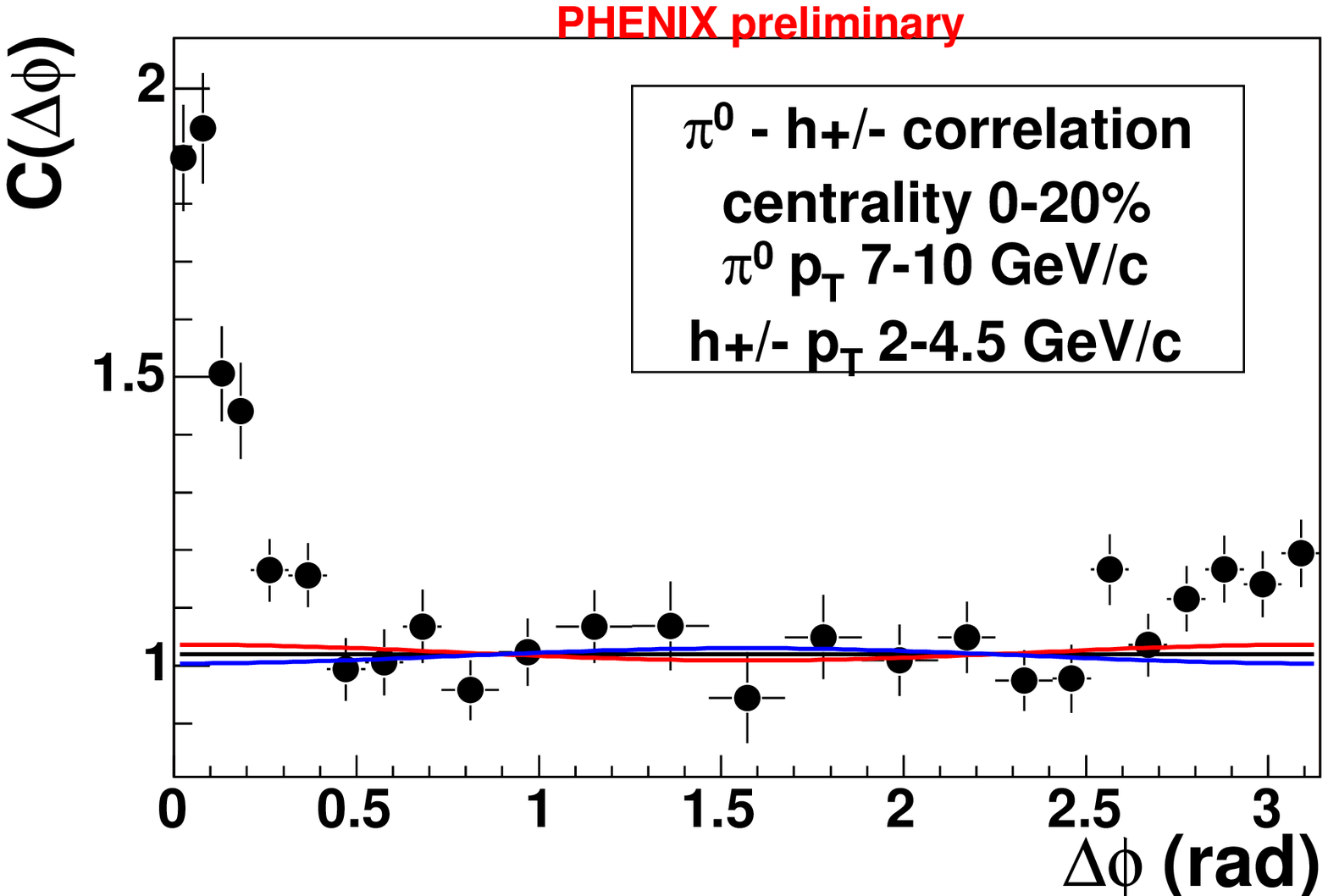}
\caption{$\pi^{0}-h$ correlations in Cu+Cu (top) and Au+Au
(bottom) collisions.}\label{fig:AACorr}
\end{center}
\end{minipage}
\end{figure}

Due to intrinsic $k_{T}$ and hard and soft gluon radiation,
parton-parton collisions are not necessarily collinear in the
nucleon-nucleon center-of-mass frame. As a result, the outgoing
di-jets are acoplanar, not back-to-back. Acoplanarity of the
di-jets can be determined by the angle between the jets. Assuming
independent fragmentation one can write $\Delta\phi_{F} =
\Delta\phi_{N} + \phi_{jj}$, where $\Delta\phi_{F}$ is the angle
between the fragments of the di-jets, $\Delta\phi_{N}$ is the
angle between the fragments of the jets, and $\phi_{jj}$ is the
angle between the di-jets. These three angles are statistically
independent and so one can find
\begin{equation}\label{eqn:sinsq}
\left<\rm{sin}^{2}\phi_{jj}\right> =
\frac{\left<\rm{sin}^{2}\Delta\phi_{F}\right> -
\left<\rm{sin}^{2}\Delta\phi_{N}\right>}{1 -
2\left<\rm{sin}^{2}\Delta\phi_{N}\right>}
\end{equation}
The terms on the right side are fixed by the RMS widths of the
near- and away-side correlations.

In d+Au collisions multiple scattering, which is responsible for
Cronin enhancement~\cite{dAuEnhance}, should increase this
acoplanarity of the di-jets. Fig.~\ref{fig:sinsq} shows the
difference of $\rm{sin}^{2}\phi_{jj}$ in d+Au and p+p collisions
as a function of the trigger $p_{T}$ where the trigger is either a
charged (squares) or neutral (circles) pion.  These data are
consistent with no difference between p+p and d+Au, which
constrains the amount of di-jet broadening from the cold nuclear
medium at RHIC.

\section{Jets in A+A Collisions}
\begin{figure}[tb]
\begin{minipage}[b]{80mm}
\begin{center}
\includegraphics[width=70mm,height=70mm]{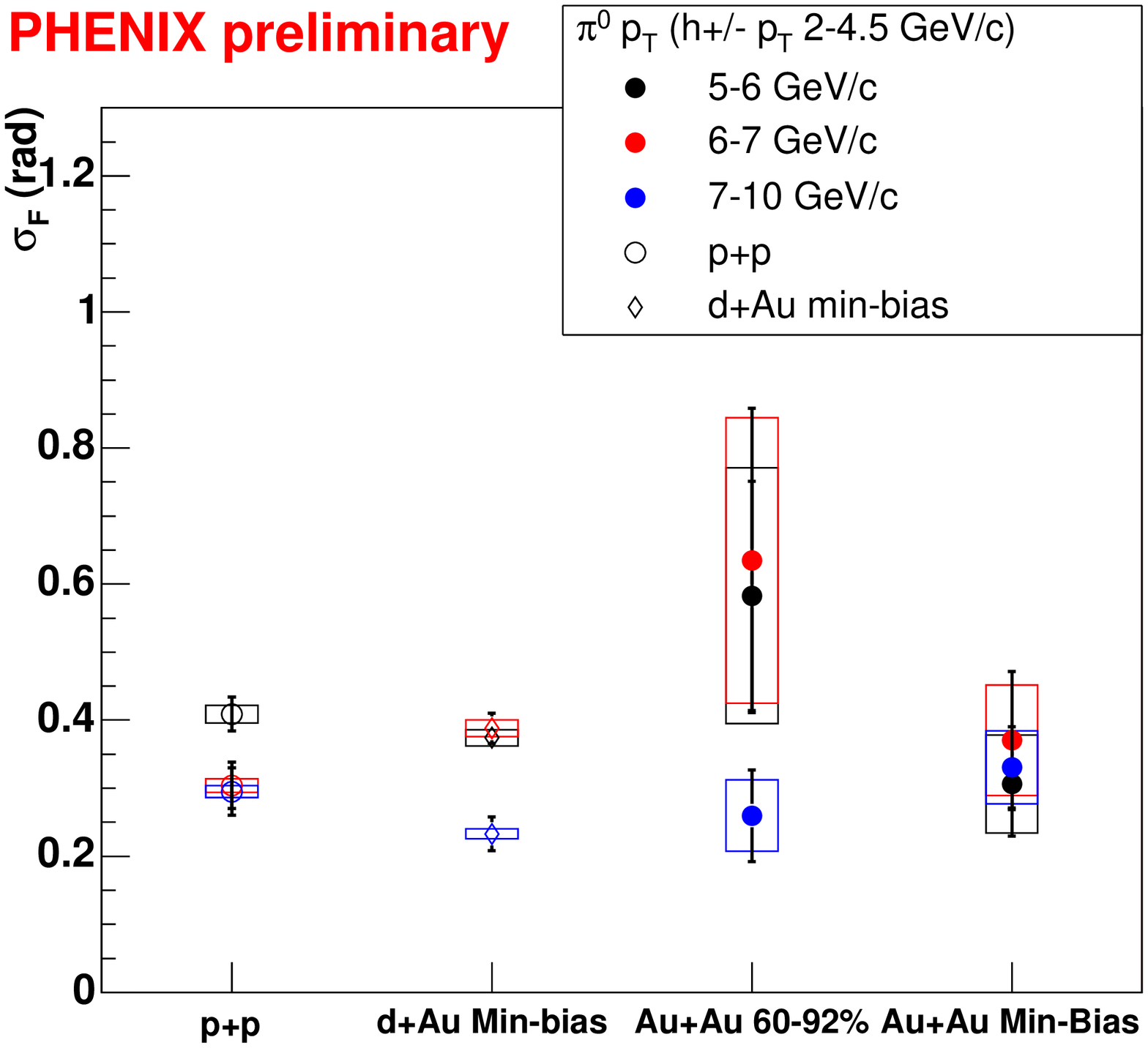}
\caption{Away-side widths of the high-$p_{T}$ jet distribution in
Au+Au collisions.}\label{fig:AAwidths}
\end{center}
\end{minipage}
\hspace{\fill}
\begin{minipage}[b]{80mm}
\begin{center}
\includegraphics[width=80mm,height=60mm]{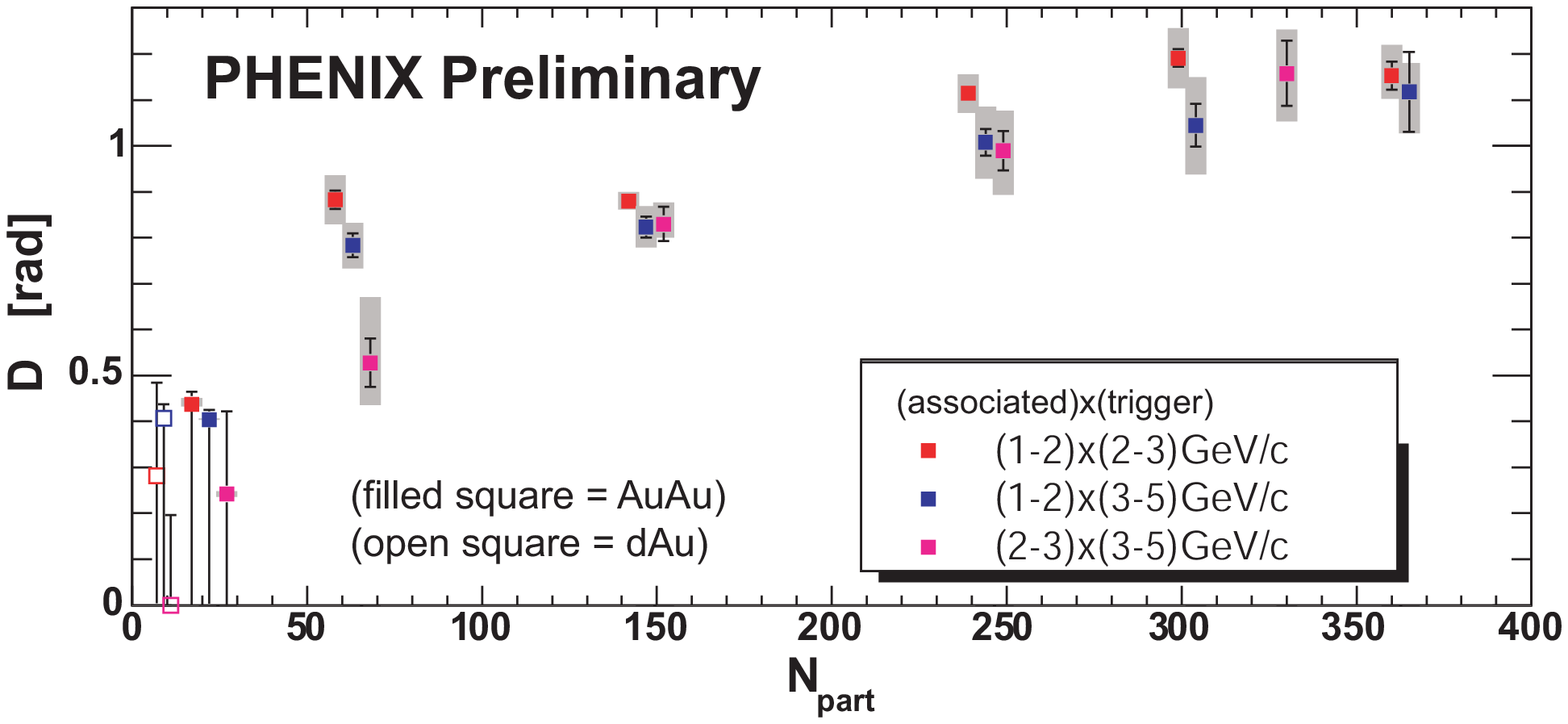}
\caption{$N_{part}$ dependence of splitting angle from Au+Au
correlations.}\label{fig:AuAuSplit}
\end{center}
\end{minipage}
\end{figure}

\begin{figure}[tb]
\begin{minipage}[b]{80mm}
\begin{center}
\includegraphics[width=80mm,height=60mm]{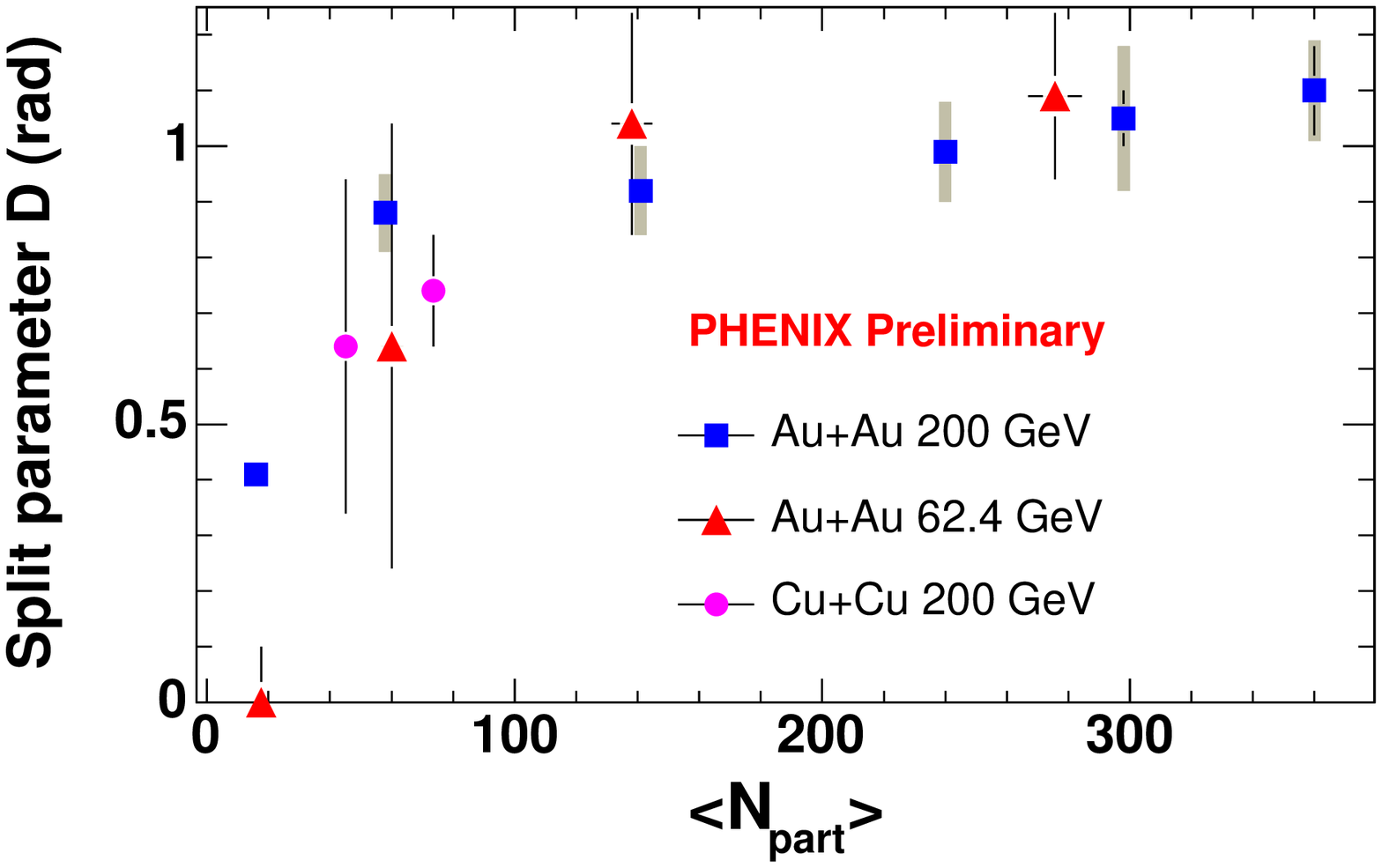}
\caption{Energy and system-size dependence of the splitting
angle.}\label{fig:SysEnerSplit}
\end{center}
\end{minipage}
\hspace{\fill}
\begin{minipage}[b]{80mm}
\begin{center}
\includegraphics[width=80mm,height=60mm]{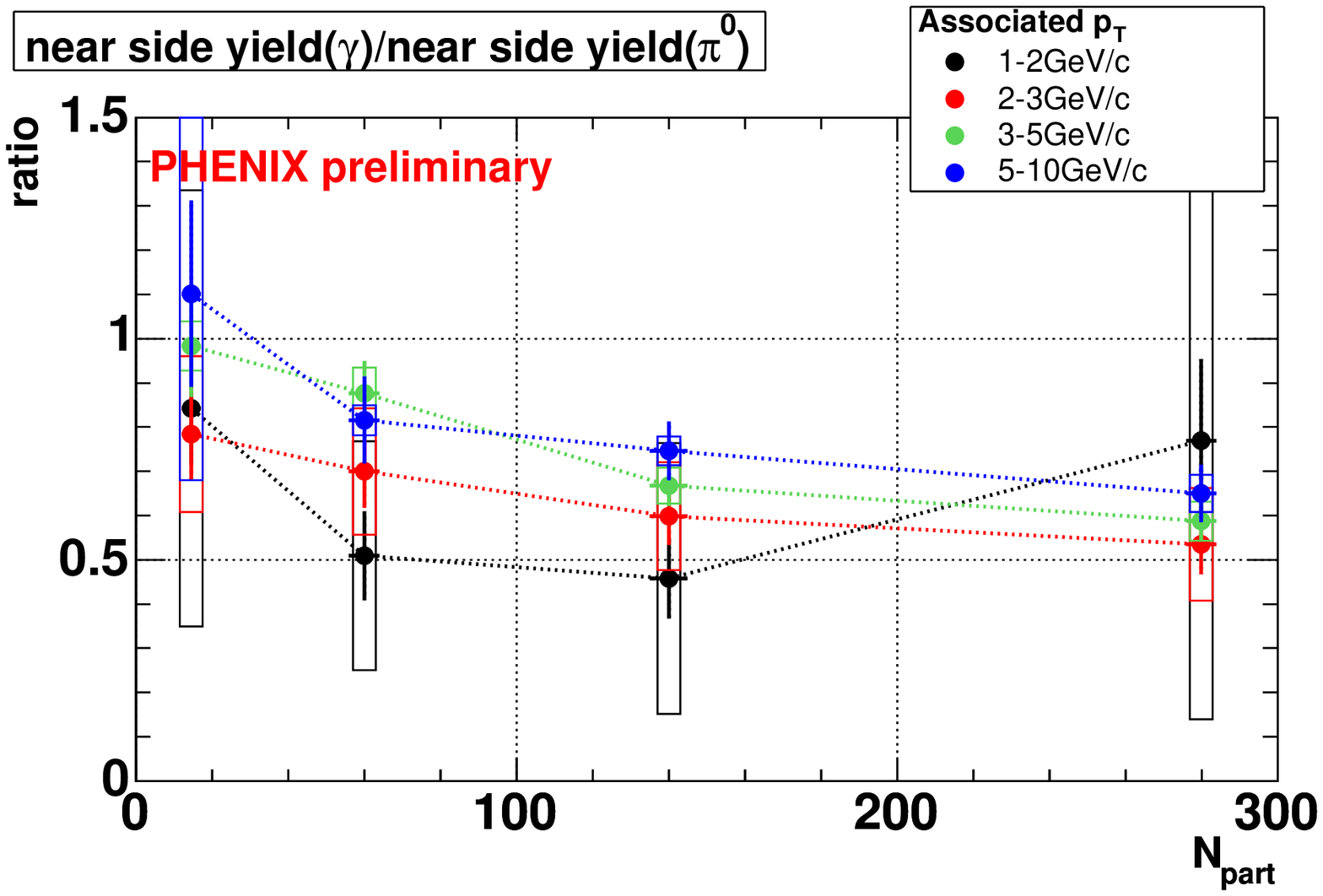}
\caption{Ratio of near angle yields with inclusive $\gamma$
triggers to $\pi^{0}$ triggers.}\label{fig:gammaToPi0YieldRatio}
\end{center}
\end{minipage}
\end{figure}

With the recent high-statistics Cu+Cu and Au+Au runs, data is
available to extended jet correlations to higher $p_{T}$ ($>$ 5
GeV) than previously measured.  The upper panel of
Fig.~\ref{fig:AACorr} shows an example of a Cu+Cu $\pi^{0}$-h
correlations in the 10\% most central events. These correlations
are normalized so that the vertical scale is the
signal-to-background of the jets. There is a large S/B and the
near- and away-side jet distributions are observed. The lower
panel of Fig.~\ref{fig:AACorr} also shows high-$p_{T}$ $\pi^{0}$-h
correlations in Au+Au collisions for the 20\% most central
collisions where the away-side jet observable at 20\% S/B. We
extract the away-side widths and compare them to the widths from
p+p and d+Au collisions (Fig.~\ref{fig:AAwidths}). The widths are
consistent between p+p and Au+Au for this $p_{T,trig}$ range.

At lower momentum ($p_{Ttrig} <$ 5 GeV) the jet shape becomes
modified such that a local minimum appears at $\Delta\phi$=$\pi$.
This has spurred theoretical interest in Mach cones or
Cerenkov-like radiation~\cite{MachCerenkovCones} as possible
physical mechanisms which could produce such a structure.
Experimentally we study this shape quantitatively by
parameterizing the away-side shape as two Gaussians with similar
width but offset symmetrically about $\Delta\phi$=$\pi$ by a
splitting angle D.
\begin{equation}
C_{away}\left(\Delta\phi\right) =
\rm{exp}\left(\frac{-\left(\Delta\phi-\pi-D\right)^{2}}{2w^{2}}\right)
+
\rm{exp}\left(\frac{-\left(\Delta\phi-\pi+D\right)^{2}}{2w^{2}}\right)
\end{equation}
Fig.~\ref{fig:AuAuSplit} shows the angle D as a function of
$N_{part}$ for different low-$p_{T}$ h-h correlations.  There is a
smooth increase of D at peripheral collisions to an apparent
saturation at central collisions.  We also observe that the lowest
$p_{T}$ combination is systematically higher than the two higher
$p_{T}$ ranges.  Low-$p_{T}$ correlations have also been performed
in Cu+Cu at $\sqrt{s}$ = 200 GeV and Au+Au at $\sqrt{s}$ = 62 GeV.
These correlations exhibit the same away side shape structure as
seen in Au+Au. Fig.~\ref{fig:SysEnerSplit} shows the same D
parameter as a function of $N_{part}$ for the different collision
systems.  All of the systems follow $N_{part}$ scaling and, within
errors, the angle D is independent of colliding energy.

An ideal probe for the jet modification in medium is the use of
prompt $\gamma$-h correlations.  PHENIX has shown a factor of 2
excess of photons above the hadronic decay background above 5
GeV/c in the most central Au+Au collisions consistent with prompt
$\gamma$ production as calculated by pQCD~\cite{DirectGammaAA}. We
have analyzed inclusive $\gamma$-h correlations.  The near angle
region of these correlations are populated by decay and prompt
$\gamma$ triggers.  Fig.~\ref{fig:gammaToPi0YieldRatio} shows the
associated particle yield per trigger in the near-side from
inclusive $\gamma$ triggers compared to $\pi^{0}$ triggers. One
sees a decrease in the ratio with centrality consistent with the
fraction of prompt-$\gamma$ increasing at the higher centralities.

PHENIX has measured jet correlations in p+p, d+Au, Cu+Cu, and
Au+Au at low- and high-$p_{T}$.  Little difference between p+p and
d+Au di-jet acoplanarity is seen. We observe in Cu+Cu and Au+Au
collisions an away-side jet widths that are similar to p+p. At
lower-$p_{T}$ a strong away-side modification is seen and the
shape seems to scale with $N_{part}$. Finally, we have shown
evidence that in central Au+Au collisions we are sensitive to
prompt $\gamma$-h correlations.

\end{document}